\documentclass[twocolumn]{article}
\usepackage{lipsum}
\usepackage{todonotes}
\usepackage{amsmath}
\usepackage{placeins}
\usepackage{booktabs}
\newcommand{\ra}[1]{\renewcommand{\arraystretch}{#1}}
\usepackage{color}
\usepackage{hyperref}
\usepackage[margin=1.8cm]{geometry}		 			% sets margin
\usepackage[font=small,labelfont=bf]{caption}		% sets caption headers bold and font small
\usepackage{enumitem}

\usepackage{amssymb}
\usepackage{pifont}
\newcommand{\cmark}{\ding{51}}%
\newcommand{\xmark}{\ding{55}}%
\usepackage{mathtools}

\usepackage{gensymb}

\graphicspath{Figures}

\usepackage{etoolbox}
\title{Battery health prediction under generalized conditions using a Gaussian process transition model}
\author{Robert R. Richardson, Michael A. Osborne and David A. Howey}

\begin{document}

\maketitle
\begin{abstract}
	Accurately predicting the future health of batteries is necessary to ensure reliable operation, minimise maintenance costs, and calculate the value of energy storage investments.
	The complex nature of degradation renders data-driven approaches a promising alternative to mechanistic modelling.
	This study predicts the changes in battery capacity over time using a Bayesian non-parametric approach based on Gaussian process regression. These changes can be integrated against an arbitrary input sequence to predict capacity fade in a variety of usage scenarios, forming a generalised health model. 
	The approach naturally incorporates varying current, voltage and temperature inputs, crucial for enabling real world application.
	A key innovation is the feature selection step, where arbitrary length current, voltage and temperature measurement vectors are mapped to fixed size feature vectors, enabling them to be efficiently used as exogenous variables.
	The approach is demonstrated on the open-source NASA Randomised Battery Usage Dataset, with data of 26 cells aged under randomized operational conditions. Using half of the cells for training, and half for validation, the method is shown to accurately predict non-linear capacity fade, with a best case normalised root mean square error of 4.3\%, including accurate estimation of prediction uncertainty.  % 
\end{abstract}

%\subsection*{Graphical abstract}
%\begin{figure*}[!h]
%	\centering
%	{\includegraphics[width=0.62\textwidth]{\string"Figures/GraphicalAbstract\string".pdf}}
%	\caption{Graphical abstract}
%\end{figure*}	

\subsection*{Keywords}
Gaussian process regression, lithium-ion, battery, degradation, prognostics, health

\subsection*{Highlights}
\begin{itemize}
	\item Gaussian process transition model of battery degradation
	\item Predicts future capacity and associated uncertainty
	\item Arbitrary length current, voltage and temperature data mapped to fixed size input vectors
	\item Demonstrated on dataset of 26 cells under randomized usage 
	\item Best case normalised root mean square error of 4.3\%
\end{itemize}

	\section{Introduction}
	
	%\head{Motivation}
	Electrochemical batteries, such as lithium-ion and lead-acid cells, experience degradation over time and during usage, leading to decreased energy storage capacity and increased internal resistance. Being able to predict the rate of degradation and the remaining useful life (RUL) of a battery is important for performance and economic reasons. For example, in an electric vehicle, the driveable range is directly related to the battery capacity. For energy storage asset valuation, depreciation, warranty, insurance and preventative maintenance purposes, predicting RUL at design stage and during operation is crucial, and the investment case is strongly dependent on the degradation behaviour \cite{wankmueller2017impact}. To estimate accurately the second hand value of assets such as EVs and grid batteries, credible predictions of RUL are required.
	
	Unfortunately, battery degradation is caused by many complex interacting chemical and mechanical processes~\cite{birkl2017degradation, ruetschi2004aging}, and physical modelling from first principles is very challenging. To mitigate uncertainty in lifetime, batteries are often over-sized and under-used, which results in increased system costs and sub-optimal performance. Hence, new approaches for accurate health prognostics are required, and form an important component of a modern battery management system or energy management system.

	%\head{Battery degradation}
	Since the performance of a battery in an application is largely dependent on its nominal capacity and internal resistance, the state of health (SoH) is typically defined by one or both of these parameters. In the present case we consider just cell capacity as the SoH metric, but the methods outlined in this paper could be applied to any other SoH metric, such as internal resistance, or capacity at some nominal C-rate. A variety of techniques may be applied for SoH measurement and estimation \cite{farmann2015critical}, but in this paper we simply assume that SoH metrics are available, for example from a battery management system.

	%\head{Conventional approaches - empirical curve fit}
    The conventional approach to battery SoH forecasting is to fit a parametric function to a broad set of ageing data measured under controlled laboratory conditions. Careful judgement is required to decide on the exact form of parametric model to use. For example, Schimpe \cite{schimpe2018comprehensive} investigated both calendar and cycle ageing of lithium iron phosphate (LFP) batteries with respect to temperature and state of charge (SoC) and found that capacity evolved with time according to
    \begin{equation}
        Q_{\text{loss}}=k_1 \sqrt{t} + k_2 \sqrt{Q_{\text{tot}}} + k_3  \sqrt{Q_{\text{ch}}} + k_4  Q_{\text{ch}},
        \label{eqn:schimpe}
    \end{equation}
    where $Q_{\text{loss}}$ is the capacity fade at some point in time, $k_{1...4}$ are empirically fitted stress factors that are a function of temperature, charging current, time and SoC, $Q_{\text{tot}}$ is the total charge throughput to time $t$, and  $Q_{\text{ch}}$ is the charge throughput only during charging, to time $t$. The stress factors $k_{1...4}$ typically fit an Arrhenius equation of the form
    \begin{equation}
            k = k_{\text{ref}} \text{ exp}\left[ \alpha \left( u - u_{\text{ref}} \right) \right],
    \end{equation}
    where $\alpha$ is a fitted constant, $u$ is some input such as current or the reciprocal of temperature, and $u_{\text{ref}}$ is reference value for that input. Very similar approaches have been developed by others for LFP batteries \cite{wang2011cycle}, and for a variety of other chemistries including NMC lithium-ion \cite{schmalstieg2014holistic, ecker2012development} and lead-acid \cite{dufo2014comparison}. These empirical degradation models are essentially parametric curve fitting using specified underlying functions such as exponentials, square roots etc. For some kinds of battery degradation data, such as \cite{birkl_howey_data}, these approaches may give a reasonable fit to the measured behaviour, although there is very little information in the literature about their long term predictive accuracy. These approaches also require the form of the model to be specified \textit{a priori}, for example (\ref{eqn:schimpe}) assumes decoupling of inputs, and this may not be the case. Additionally, many degradation datasets exhibit an accelerated capacity fade regime in later life (see \cite{harris2017failure}), and this approach is not able to model such a regime change. Also, accuracy may be limited when environmental and load conditions differ from the training dataset. %, and when the capacity fade depends on additional contributions from unknown sources. % but same for ML!
    
    % bottom up models
    As an alternative approach to empirical parametric functions fitted to laboratory test data, others have developed `first principles' electrochemical models of battery ageing. These propose and model a set of underlying physical ageing mechanisms. For example a popular ageing mechanism is growth of the anode solid electrolyte interphase (SEI) through reduction of the ethylene carbonate in the electrolyte, modelled as a diffusion-limited single step charge transfer reaction \cite{kupper2017multi, pinson2013theory}. This can be augmented to include additional physics related to lithium plating \cite{Yang2017}, particle cracking \cite{Deshpande2017} and other mechanisms. Although reasonable results are demonstrated for calendar ageing, huge challenges remain with respect to parametrisation and validation of such models, and what physics to include to capture all the relevant ageing mechanisms and their interactions.
    
	%\head{Data-driven approaches}
	In contrast to these approaches, so-called data-driven battery ageing models are beginning to be investigated. These have some similarities with the empirically fitted functions previously discussed, but new techniques from machine learning allow much greater flexibility in these models than can be obtained using pre-specified parametric functions.  The simplest formulation of this is direct fitting of capacity data with respect to time, or cycle count, which allows RUL estimation by extrapolation to future values. A variety of data-driven techniques have been explored in this context, including non-parametric approaches such as support vector machines ~\cite{hu2016battery,patil2015novel,wang2013prognostics,nuhic2013health}, and Bayesian non-parametric approaches such as Gaussian process (GP) regression~\cite{goebel2008prognostics,saha2008uncertainty,he2011prognostics}. A non-parametric model is one whose expressivity (as would increase with the degree of a polynomial, for instance) naturally adapts to the complexity of data. Rather than having no parameters, a non-parametric model is perhaps better thought of as one with a number of parameters that can scale with the data and could become arbitrarily large. Bayesian approaches naturally incorporate estimates of uncertainty into predictions, allowing a model to acknowledge the varying probabilities of a range of possible future health values, rather than just giving a single predicted value.
	
	These approaches have been demonstrated to work well when a battery health dataset is available for batteries that have all been cycled in a similar way. For example, our previous work \cite{richardson2017gaussian} on RUL prediction applied a multiple-output Gaussian process model to incorporate data from multiple batteries, all cycled in the same way, demonstrating a large improvement in accuracy of RUL estimation over existing methods. However, for real world RUL prediction at design stage, or for preventative maintenance, a much more flexible approach is needed that allows health predictions to be made as a function of the changing stress factors such as time, charge throughput and temperature etc. The previously discussed parametric models can incorporate dependence on external inputs, but are limited to pre-specified functions. In other words, they assume that the shape of the degradation trajectory is known \emph{a priori}, which limits their applicability.
	
    To address this, we introduce the idea of a Bayesian non-parametric transition model for battery health. Rather than fitting the SoH data directly as a function of time or cycle count, the model predicts the \emph{changes} in SoH from one point to the next as the battery is used, as a function of the usage. This is explained in detail in the next section.

\FloatBarrier
\section{Method}

	\subsection{Transition model}
The approach in this paper formulates a transition model to predict the capacity changes between periods of usage that we term `load patterns'. We define this differently to a standard battery charge-discharge cycle, instead it is the time-series of current, voltage and temperature data between any two capacity measurements or estimates, $Q_{i}$ and $Q_{i+1}$. Load patterns do not need to be uniformly spaced, i.e.\ they could be short or long periods of usage, and might include multiple charge-discharge events.

	The goal of a regression problem is to learn the mapping from input vectors $\textbf{x}$ to outputs $y$, given a labelled training set of input-output pairs $\mathcal{D} = \{(\textbf{x}_i, y_i)\}_{i=1}^{N_D}$, where $N_D$ is the number of training examples.
	In the present case, the inputs ${\textbf{x}}_i \in \mathbb{R}^{+}$ are vectors of selected features (see section \ref{sec:featex}) for load pattern $u_i$, and the outputs ${y}_i \in \mathbb{R}^{+}$ are the corresponding differences in measured capacity between load pattern $u_i$ and $u_{i+1}$.
	The underlying model takes the form ${y} = f({\textbf{x}}) + \varepsilon$, where $f({\textbf{x}})$ represents a latent function and $\varepsilon \sim \mathcal{N}(0, \sigma^2)$ is an independent and identically distributed noise contribution.

	The learned model can then be used to make predictions on a set of test inputs $\mathbf{x^*} = \{\textbf{x}_i^*\}_{i=1}^{N_T}$ (i.e.\ load patterns where we wish to estimate the capacity), producing outputs $\mathbf{y^*} = \{y_i^*\}_{i=1}^{N_T}$, where $N_T$ is the number of test indices.
	In our case we are interested in predicting the capacity changes in a new -- previously unseen -- battery cell, which has been exposed to a known test regime. This is called the validation or test dataset.

	\subsection{Input feature extraction}
	\label{sec:featex}
    Each load pattern, $u_i$, may contain within it an arbitrary number of \textit{time steps}, $N_i$. However, in order to use the inputs in our model, since the capacity measurements are only known per load pattern, we must first map time-series data to a fixed size input vector. In other words, assuming there are $N_i$ time-steps within a load pattern $u_i$, then the measurements {$I\in\mathbb{R}^{N_i}$, $V\in\mathbb{R}^{N_i}$, $T\in\mathbb{R}^{N_i}$} are mapped to a single $n$-dimensional input vector, $\textbf{x}$, where $n$ is the number of features of interest. Irrespective of the number of time steps in a load pattern, the size of the input vector $\textbf{x}$ is the same.
	
	For each load pattern, $u_i$, the features to be extracted are defined by prior assumptions about what causes a battery to age. As discussed in the preceding section, there are many possible different stress factors that affect battery ageing, depending on the dataset and model. However, in the dataset used it was found that accurate results could be obtained with only a small number of factors (see table \ref{tab:results}), as follows:

	The first component of the input vector, for the $i$th load pattern, is the total time elapsed during the load pattern, given by
	$${x}_{i,1} = \Delta t = t_{i+1} - t_{i},$$
		where $t_{i}$ and $t_{i+1}$ are the times at the start and end of the load pattern respectively.
   
   The second component is the charge throughput, $Q_{\text{thru}}$, during the load pattern, i.e.\ the total absolute current through the cell during the load pattern, given by
		$${x}_{i,2} = Q_{\text{thru}}=\int_{t_i}^{t_{i+1}} |I| \: \mathrm{d}t.$$
		
   The third component is the absolute time value, in seconds, since the beginning of the whole dataset,
   $$x_{i,3} = t.$$
   
   As discussed later, for the dataset considered here, it was found that the choice of model and number of overlapping load patterns were generally more important for determining predictive accuracy than the inclusion of additional input features. However, with a larger dataset, additional features could improve predictive accuracy. These might include the following: 
   
   Firstly, the present cell capacity,  
	$${x}_{i,4} = Q_{i}.$$
   
   Secondly, the \textit{time elapsed during which certain conditions are met}. This is achieved by defining a selection of current, voltage and temperature ranges, and evaluating the time spent by the battery within these ranges:
		$${x}_{i,j} = t_{P_l<P<P_u},$$
		for $j \in \{5,6,\dots\}$, where $P$, $P_l$ and $P_u$ are the parameters of interest, and their upper and lower bounds respectively. 
		For example, a battery's aging behaviour is expected to be affected by high or low temperatures \cite{schimpe2018comprehensive}. Hence, one might define the duration of time the battery spends (1) below 0 $^\circ$C, (2) between 0 and 40 $^\circ$C, and (3) above 40 $^\circ$C as three distinct inputs:
		$${x}_{i,5} = t_{T<0 ^\circ \text{C}}$$
		$${x}_{i,6} = t_{0  ^\circ \text{C} <T< 40 ^\circ \text{C}}$$
		$${x}_{i,7} = t_{40 ^\circ \text{C}<T}$$

	An example of an input vector for a single load pattern is given in Table \ref{tab:input-vector}. In this case, inputs were defined for ranges of temperature and current. Of course, additional inputs could also be defined by voltage ranges, but these have been omitted here for clarity of presentation. Note that the sum of all the times spent in each parameter range (e.g.\ in each temperature or current range) must equal the total time elapsed within that load pattern.

	\begin{table*}
		\centering
		\small
		\ra{1.1}
		\begin{tabular}{ccccccccccc}
			\toprule
			\multicolumn{10}{c}{$\textbf{X}$} & $\textbf{y}$  \\
			\midrule
			Capacity & \multicolumn{9}{c}{Inputs} & Output \\
			$Q [\text{Ah}]$ & $\Delta t$ & $t$ & $q_{\text{thru}} [\text{Ah}]$ & 
			$t_{T<5^\circ \text{C}}$ & $t_{5^\circ \text{C}<T<40^\circ \text{C}}$ & $t_{T>40^\circ \text{C}}$ & 
			$t_{I<2 \text{A}}$ & $t_{2\text{A}<I<3 \text{A}}$ & $t_{I>3 \text{A}}$ &  $\Delta Q [\text{Ah}]$ \\
			$2.1$ & $5.5$ & $10.6$ & $1.4$ & $0$ & $5$ & $0.5$ & $4$ & $1.25$ & $0.25$ & $-0.05$ \\
			\bottomrule
		\end{tabular}
		\caption{Example input format for a single step. Note that the values of the inputs shown here are arbitrary and for explanatory purposes only. All units are in hours, except where denoted otherwise.}
		\label{tab:input-vector}
	\end{table*}

	\subsection{Example data}
\begin{figure}[hbt]
		\centering
		{\includegraphics[width=1.0\columnwidth]{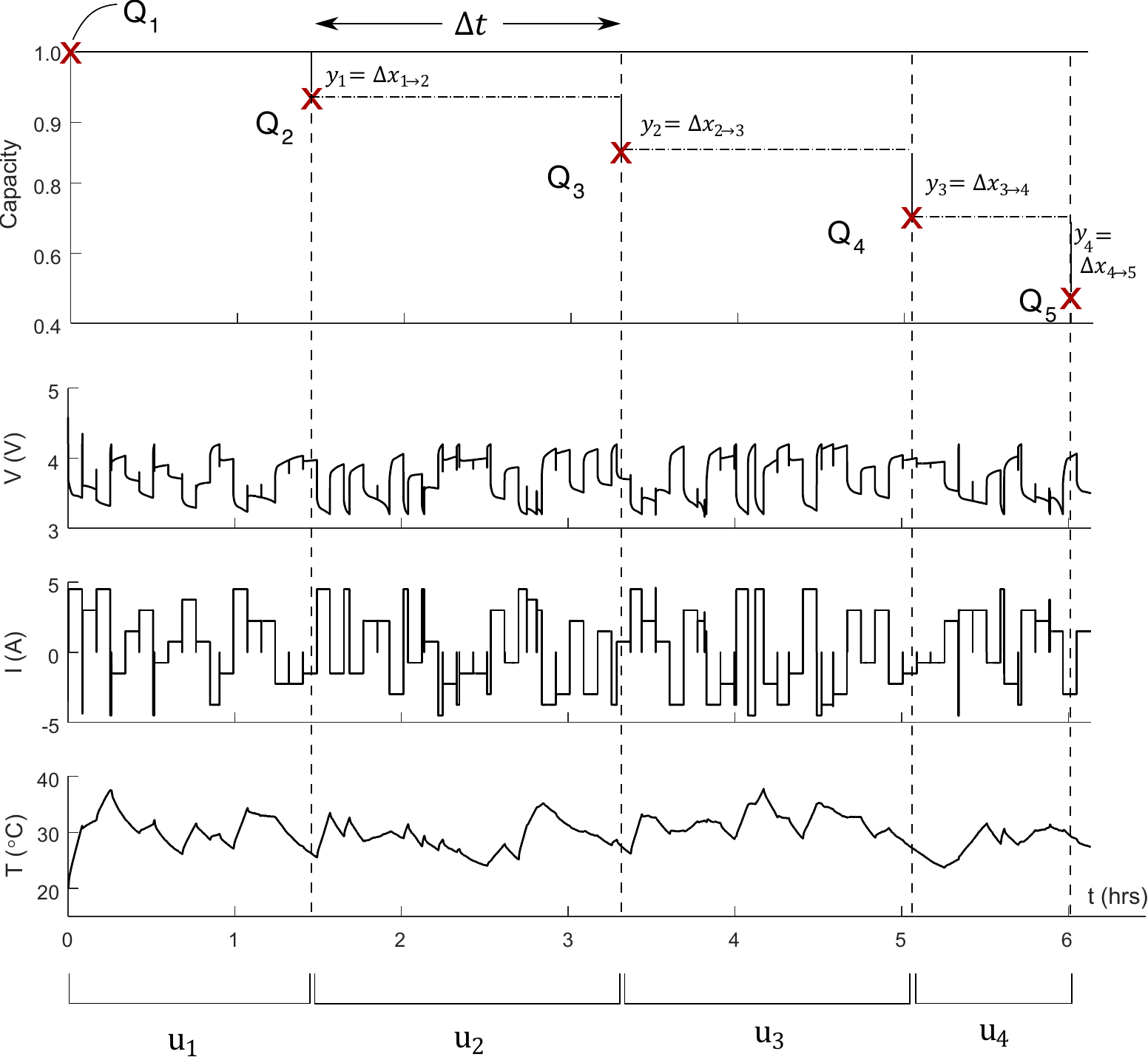}}
		\caption{Schematic diagram showing raw input data and outputs for the first four load patterns for a single cell.}
		\label{fig:overview}
	\end{figure}

    Fig.\ \ref{fig:overview} shows an exemplary schematic of the first 4 load patterns for a single cell. There is one capacity measurement ($Q_1$) at the very start of the cell's life and then 4 subsequent measurements ($Q_2 - Q_5$) at later times. The load patterns consist of everything that occurs between each capacity measurement; each load pattern is translated into equal sized input vectors, $\textbf{x}_i$.
	
\begin{figure*}[hbt]
		\centering
		\includegraphics[width=0.99\textwidth]{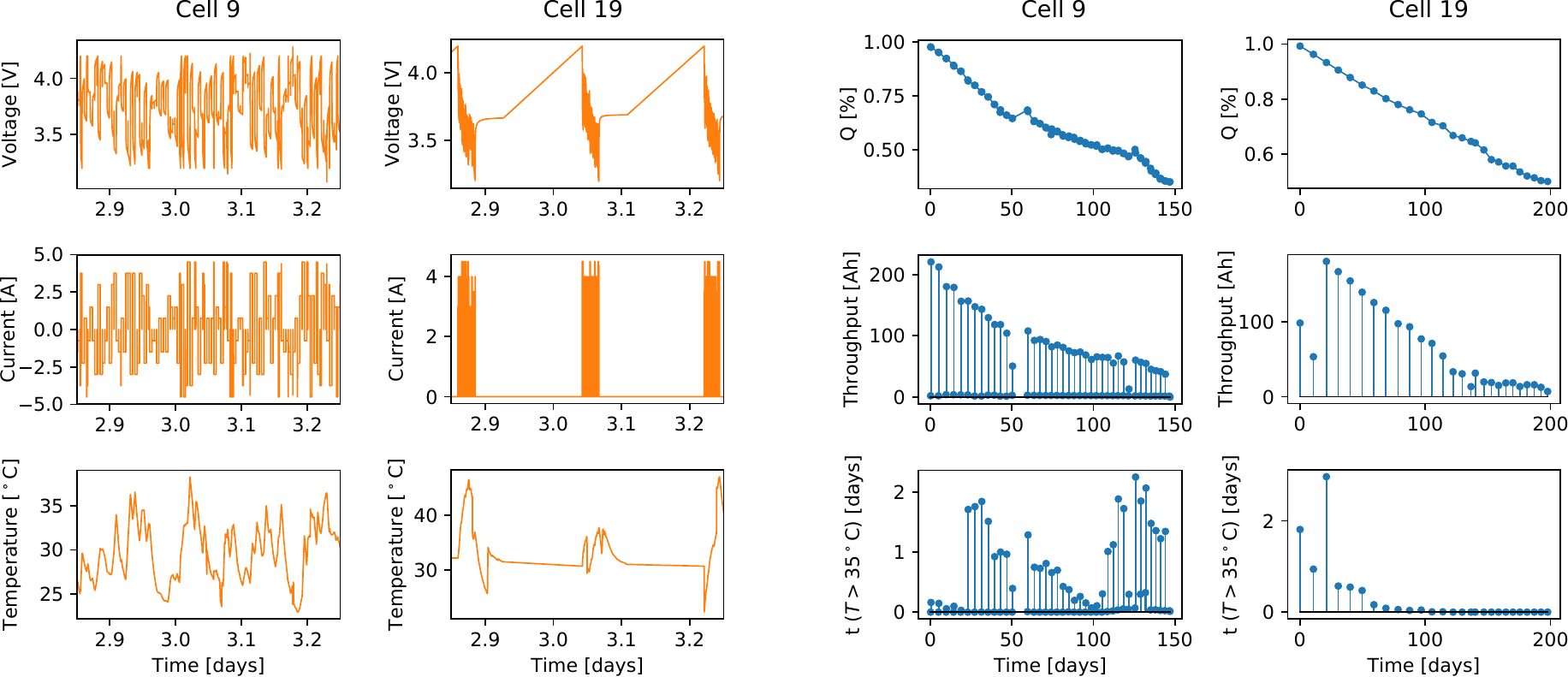}
		\caption{Examples of raw data (left two columns) and extracted input features (right two columns) for two different cells}.
		\label{fig:inputs}
	\end{figure*}

	Fig.\ \ref{fig:inputs} gives examples of real measurement data for two different cells, including the capacity values $Q$, some of the extracted inputs for each load pattern, and exemplary time series measurements corresponding to a portion of a load pattern. The dataset used for this work is explained in section \ref{sec:Datasets}.

	\subsection{Evaluation}
	The model predictions are evaluated using three different metrics, which reflect the quantities of interest in a practical application. The first is the root-mean-squared error (RMSE) in the mean output of the model (i.e.\ the capacity differences), defined as 
	\begin{equation}
	\text{RMSE}_{\Delta Q}(\hat{y_i}^*, y_i^*) = \sqrt{\frac{1}{N_T}\sum_{i=1}^{N_T}\left(\hat{y_i}^* - y_i^*\right)^2},
	\end{equation}
	where $N_T$ is the number of points to be evaluated (i.e.\ all points in the test dataset), $y_i^*$ is the measured capacity difference using the test dataset and $\hat{y}_i^*$ is the estimated mean capacity difference predicted by the model, each between load pattern $u_i$ and $u_{i+1}$.
	The second is the RMSE in actual capacity, defined as
	\begin{equation}
	\text{RMSE}_Q(\hat{Q_i}^*, Q_i^*) = \sqrt{\frac{1}{N_T}\sum_{i=1}^{N_T}\left(\hat{Q_i}^* - Q_i^*\right)^2},
	\end{equation}
	where $Q_i^*$ is the measured capacity (using the test dataset) and $\hat{Q}_i^*$ is the estimated mean capacity, each at load pattern $i$. This may also be expressed as a normalised value, to facilitate comparison with other studies, whereby the absolute capacities may be of different magnitudes:
	\begin{equation}
	\text{RMSE}_{Q, norm}(\hat{Q_i}^*, Q_i^*) = \sqrt{\left(\frac{1}{N_T}\sum_{i=1}^{N_T} \frac{(\left(\hat{Q_i}^* - Q_i^*\right)}{Q_i}\right)^2}.
	\end{equation}
	Note that it is possible for a model to perform well in one of these metrics but poorly in the other. For instance, if a model over-predicts $\Delta Q$ every second load pattern but under-predicts on alternate load patterns, the overall capacity evolution may be accurate (implying good $\text{RMSE}_{Q}$), but the individual predictions might not be (implying poor $\text{RMSE}_{\Delta Q}$).
	Hence, a good model should have low values of both these metrics.
	
	Thirdly, since the approach used here is probabilistic, the accuracy of the uncertainty estimates can also be quantified using the calibration score (CS). This is defined as the frequency of measured results in the test dataset that are within a predicted credible interval. Within a $\pm2\sigma$  interval, corresponding to a 95.4\% probability for a Gaussian distribution, the CS is given by
\begin{equation}
\text{CS}_{2\sigma} = \frac{1}{N_T}\sum_{i=1}^{N_T}\left[\lvert \hat{y_i}^* - y_i^* \rvert < 2\sigma \right].
\label{eq:calib-score}
\end{equation}

 Therefore, $\text{CS}_{2\sigma}$ should be approximately 0.954 if the uncertainty predictions are accurate, using the techniques outlined in this paper. Higher or lower scores indicate under- or over-confidence, respectively.
	
	\subsection{Gaussian process regression}
	
	This section gives a brief overview of Gaussian process regression, the main approach chosen in this paper for modelling the transition in health from one load pattern to the next.
	A Gaussian process (GP)~\cite{rasmussen2006gaussian} defines a probability distribution over functions, and is denoted as:
	\begin{equation}
	f(\mathbf{x}) \sim \mathcal{GP}\bigl(m(\mathbf{x}), \kappa(\mathbf{x},\mathbf{x}')\bigr),
	\end{equation}
	where $m(\mathbf{x})$ and $\kappa(\mathbf{x},\mathbf{x}')$ are the mean and covariance functions respectively, denoted by
	\begin{align}
	m(\mathbf{x}) & =
	\mathbb{E} [f(\mathbf{x})], \\
	\kappa(\mathbf{x}, \mathbf{x}') & =
	\mathbb{E} [\left(f(\mathbf{x}) - m(\mathbf{x})\right) \left(f(\mathbf{x}') - m(\mathbf{x}')\right)^T].
	\end{align}
	For any finite collection of input points, say ${X} = \mathbf{x}_1,...,\mathbf{x}_{N_D}$, this process defines a probability distribution $p\left( f(\mathbf{x}_1),...,f(\mathbf{x}_{N_D}) \right)$ that is jointly Gaussian, with some mean $\mathbf{m} (\mathbf{x})$ and covariance $\mathbf{K} (\mathbf{x})$ given by $K_{ij} = \kappa(\mathbf{x}_i,\mathbf{x}_j)$.
	
	Gaussian process regression is a way to undertake non-parametric regression with Gaussian processes.
	Rather than suggesting a parametric form for the function $f({\mathbf{x}, \phi})$ and estimating the parameters $\phi$ (as in parametric regression), we instead assume that the function $f(\mathbf{x})$ is a sample from a Gaussian process as defined above.
	
	In this work, we use the Mat\'ern covariance function:
	\begin{equation}
	\kappa_{\text{Ma}}({x}-{x}') = 
	\sigma_f^2 \frac{2^{1-\nu}}{\Gamma(\nu)}
	\left(\sqrt{2\nu}\frac{({x}-{x}')}{\rho}\right)^{\nu}
	\mathcal{R}_{\nu}\left(\sqrt{2\nu}\frac{({x}-{x}')}{\rho}\right)
	,
	\end{equation}
	with output scale $\sigma_f$, smoothness hyperparameter, $\nu=5/2$ (larger $\nu$ implies smoother functions)
	and $\mathcal{R}_{\nu}$ is the modified Bessel function.
	This kernel was chosen because it is suitable for functions with varying degrees of smoothness, although similar performance was observed using other common kernels, including the squared exponential~\cite{rasmussen2006gaussian}.
	
	A fuller discussion of various different kernels that may be used for GP regression in the context of battery health prediction is given in \cite{richardson2017gaussian}. Finally, we also compare performance against a linear kernel, since this is equivalent to Bayesian linear regression \cite{rasmussen2006gaussian}:
	\begin{equation}
	    k_{\text{lin}}(x,x') = \sigma^2_f (x-c)(x'-c),
	\end{equation}
	where $c$ is a constant defining the offset of the linear function. The mean function of the GP is commonly defined as $m(\mathbf{x})=0$, and we follow this convention here.
	
	Now, if one observes a labelled training set of input-output pairs
	$\mathcal{D} = \{(\mathbf{x}_i, {y}_i)\}_{i=1}^{N_D}$, predictions can be made at test indices ${X}^*$ by computing the conditional distribution $p(\mathbf{y}^* \vert {X}^*, {X}, \mathbf{y})$.
	This can be obtained analytically by the standard rules for conditioning  Gaussians~\cite{murphy2012machine}, and (assuming a zero mean for notational simplicity) results in a Gaussian distribution given by
	\begin{equation}
	p(\mathbf{y}^* \vert {X}^*, {X}, \mathbf{y}) = \mathcal{N}(\mathbf{y}^* \vert \mathbf{m}^*, \mathbf{\sigma}^*)
	\end{equation}
	where
	\begin{align}
	\mathbf{m}^* & = \mathbf{K}({X}, {X}^*)^T
	\mathbf{K}({X}, {X})^{-1}
	\mathbf{y}\\
	\mathbf{\sigma}^* & = \mathbf{K}({X}^*,{X}^*)
	- \mathbf{K}({X}, {X}^*)^T
	\mathbf{K}({X}, {X})^{-1}
	\mathbf{K}({X}, {X}^*).
	\end{align}
	
	The values of the covariance hyperparameters $\theta$ may be optimised by minimising the negative log marginal likelihood defined as $\text{NLML} = -\log p(\mathbf{y} \vert {X}, \theta)$.
	Minimising the NLML automatically performs a trade-off between bias and variance, and hence ameliorates over-fitting to the data \cite{bishoppattern}. Given an expression for the NLML and its derivative with respect to $\theta$ (both of which can be obtained in closed form), $\theta$ can be estimated using gradient-based optimization.
	The Python GPy library was used to implement these algorithms.

\subsection{Gradient boosting}
\label{subsec:GB}
As a state-of-the-art comparison to Gaussian process regression, we also investigated predictive performance with an alternative technique, gradient boosting. This is a popular data-driven time series modelling approach based on combining an ensemble of weak prediction models into a stronger model~\cite{murphy2012machine}. While this approach is not inherently probabilistic, and does not output a full covariance matrix for the predictions, it can be trained using quantile regression (QR) to approximately predict a probability distribution. Quantile regression deliberately introduces a bias in the prediction in order to estimate statistics. The loss function is modified such that instead of identifying the mean of the variable to be predicted, QR seeks the median and any other desired quantiles. To identify the upper and lower bounds of a prediction interval, QR is repeated at several different quantiles. One advantage of this method is that asymmetric intervals can be predicted. On the other hand, it is not clear how the confidence intervals for $Q$ should be calculated from the values for $\Delta Q$, since the full covariance matrix is unavailable. In this case, we simply centred the intervals around the mean, and fitted a Gaussian distribution in order to achieve this.
	
\FloatBarrier
\section{Dataset\label{sec:Datasets}}
\begin{figure}%[h]
	\centering
	{\includegraphics[width=0.8\columnwidth]{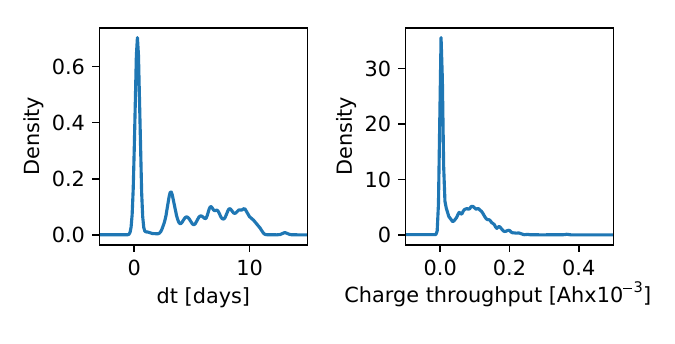}
	\includegraphics[width=1\columnwidth]{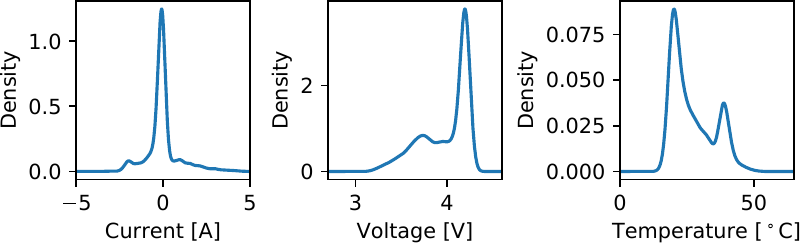}}
	\caption{Distribution of input data for each parameter.}
	\label{fig:datadist}
\end{figure}

The battery dataset used here was obtained from the NASA Ames Prognostics Center of Excellence Randomized Battery Usage
\href{https://ti.arc.nasa.gov/tech/dash/pcoe/prognostic-data-repository/}{Repository} \cite{bole2014randomized}. The data in this repository were first used in Ref.~\cite{bole2014adaptation} for an investigation into capacity fade under randomized load profiles. The data are randomised in order to better represent practical battery usage. This is ideal for training a data-driven model. Fig.~\ref{fig:datadist} gives smoothed histograms computed from the cell data showing the ranges of times, charge throughput, currents, voltages and temperatures that are explored by this dataset.

An overview of the battery dataset is given in Table~\ref{tab:Data}. The cells used have a relatively high energy density, but short lifetime. The remainder of this subsection describes the cycling and characterisation procedure, based on~\cite{bole2014randomized}. 
For this study we used data from 26 of the 28 total battery cells available in the repository (cells 16 and 17 were omitted, since these were found to contain spurious data resulting in certain cycles having negative duration).
The cells were grouped into 7 groups of 4, with each group undergoing a different randomized cycling procedure as described in Table~\ref{tab:NASA-groups}.

The first 5 groups were cycled at room temperature throughout the duration of the experiments, whilst groups 6-7 were cycled at 40~$^\circ$C.
In all cases a characterisation test was periodically carried out, whereby a 2~A charge-discharge cycle was applied (i.e.\ approximately 1C) between the cell voltage limits -- these discharge curves were used to evaluate the capacity as an indicator of state of health.
There were a total of 950 discharge curves available across all cells (i.e.\ $\sim34$ curves per cell).

\begin{figure*}%[h]
	\centering
	{\includegraphics[width=1\textwidth]{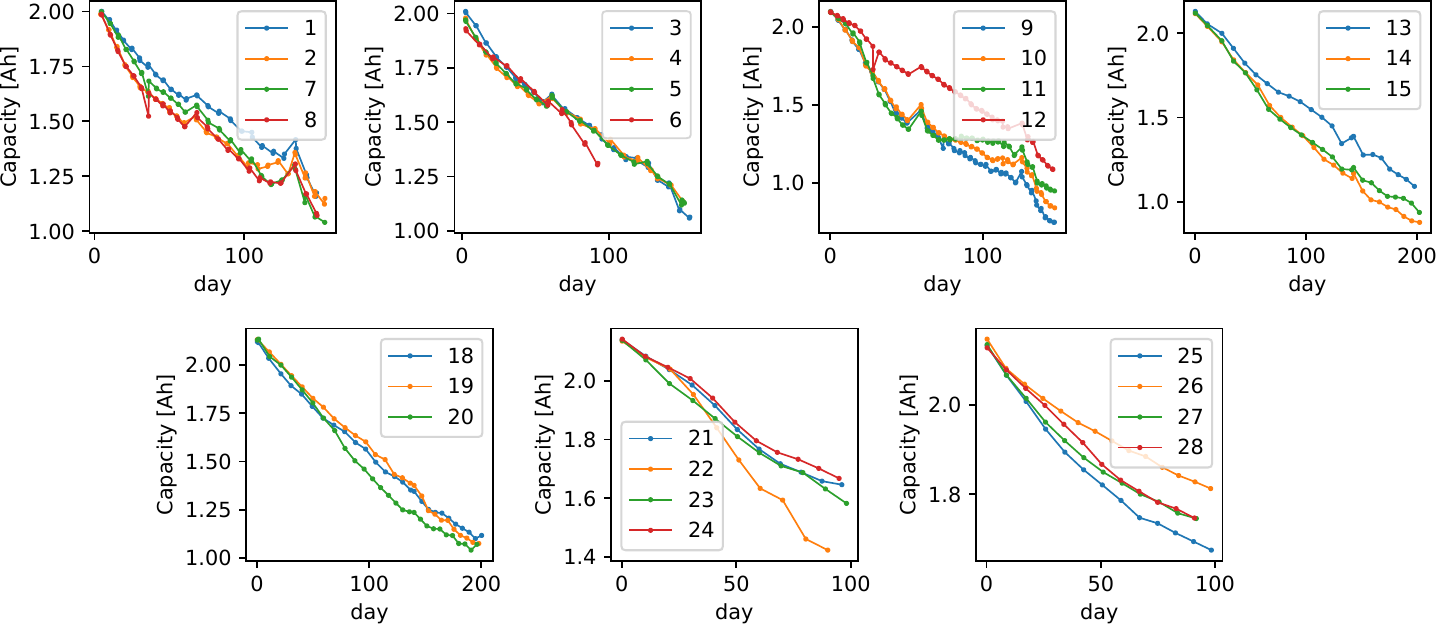}}
	\caption{Measured cell capacities for each group of similarly cycled cells.}
	\label{fig:data}
\end{figure*}

The cell capacity was calculated by integrating the current from each of the 2~A charge curves. Calculated capacities for the cells in each group are plotted against time in Fig.~\ref{fig:data}. The evolution of the capacity is quite different for each group of cells.

\begin{table}[h]
	\centering
	\small
	\ra{1.1}
	%	\begin{tabular}{@{}lll@{}}
	\begin{tabular}{ll}
		\toprule
		Manufacturer & LG Chem \\
		Form factor & 18650 \\
		Chemistry & Lithium cobalt oxide vs.\ graphite \\
		\# cells & 26 \\
		\# $Q$ samples & 950 \\
		$Q$ range (Ah) & 2.10 $\rightarrow$ 0.80 \\
		Cycling & 7 groups each with different regime \\
		\bottomrule
	\end{tabular}
	\caption{Dataset overview. Row `\# $Q$ samples' indicates total number of capacity measurements, i.e.\ approximate total number of health transitions. Row `$Q$ range' indicates values of the maximum initial capacity and minimum final capacity.}
	\label{tab:Data}
\end{table}

\begin{table}[h]
    \centering
	\small
	\ra{1.1}
	\begin{tabular}{p{0.49\textwidth}}
		\toprule
		\textbf{Group 1 (Cells 1, 2, 7, 8)}\\
		%		\midrule
		Repeatedly charged to 4.2~V using a randomly selected duration between 0.5 hours and 3 hours, then discharged to 3.2~V using a randomized sequence of discharging currents between 0.5~A and 4~A. Reference characterisation every 50 cycles.\\
		\textbf{Group 2 (Cells 3-6)}\\
		Same as group 1 except charging cycle not randomized.\\
		\textbf{Group 3 (Cells 9-12)}\\
		Operated using a sequence of charging/discharging currents between -4.5~A and 4.5~A. Each loading period lasted 5 minutes. Reference characterisation carried out after 1500 periods (about 5 days).\\
		\textbf{Group 4 (Cells 13-15)}\\
		Repeatedly charged to 4.2~V and then discharged to 3.2~V using a randomized sequence of discharging currents between 0.5~A and 5~A. A customized probability distribution skewed towards selecting higher currents was used to select a new load setpoint every 1 minute during discharging.\\
		\textbf{Group 5 (Cells 18-20)}\\
		Same as Group 4 except the probability distribution was designed to be skewed towards selecting lower currents.\\
		\textbf{Group 6 (Cells 21-24)}\\
		Same as Group 5 except with ambient temperature of 40$^\circ$C.\\
		\textbf{Group 7 (Cells 25-28)}\\
		Same as Group 4 except with ambient temperature of 40$^\circ$C.\\
		\bottomrule
	\end{tabular}
	\caption{NASA data load profiles. Each group of cells underwent a different loading procedure. Full details in~\cite{bole2014randomized}.}
	\label{tab:NASA-groups}
\end{table}

%\FloatBarrier
%\clearpage 

\section{Results}
\label{sec:results}
%\subsection{Model comparisons}
We considered 6 different configurations of data-driven transition model, as defined in Table~\ref{tab:results}, in order to show a range of comparisons in predictive accuracy. In each case, the model was trained on the data from even numbered cells (i.e.\ all the mappings between inputs and capacity drops across all of those cells), and subsequently tested on the odd numbered cells. 

\begin{table*}[ht]
	\centering
	\small
	\ra{1.1}
	\begin{tabular}{llcccccccccc}
		\toprule
        No. & Model & Kernel & Lags & \multicolumn{3}{c}{Inputs} & \multicolumn{2}{c}{dQ (Ah)} & \multicolumn{3}{c}{Q (Ah)}	\\
        & & & & $\Delta t$ (s)  & Q$_{\text{thru}}$ (Ah) & t (s) & RMSE & CS$_{2\sigma}$ & RMSE & RMSE$_{\text{norm}}$ & CS$_{2\sigma}$ \\
		\midrule
		1. & GP & Ma5 & 6 & \cmark & \cmark & \xmark & \textbf{0.0201} & 0.959 & \textbf{0.070} & \textbf{0.043} & \textbf{0.967} \\
		2. & GP & Ma5 & 1 & \cmark & \cmark & \cmark & 0.0236 & \textbf{0.950} & 0.116 & 0.086  & 0.922 \\
		3. & GP & Lin & 6 & \cmark & \cmark & \xmark & 0.0284 & 0.939 & 0.186 & 0.173 & 0.839 \\
		4. & GP & Lin & 1 & \cmark & \cmark & \cmark & 0.0319 & 0.945 & 0.642 & 0.593 & 0.241 \\
		5. & SKGB & n/a & 6 & \cmark & \cmark & \cmark & 0.0244 & 0.850 & 0.089 & 0.067 & 0.846 \\
		6. & SKGB & n/a & 1 & \cmark & \cmark & \xmark & 0.0246 & 0.889 & 0.125 & 0.106 & 0.757 \\
		\bottomrule
	\end{tabular}
	\caption{Results for the 6 different model combinations. The best values of each metric are indicated in bold.}
	\label{tab:results}
\end{table*}

Models 1 and 2 use a GP with a Mat\'ern kernel.
The difference between these two models is the way in which long term trends are captured.
For model 1, data from the preceding 6 load patterns were all used as inputs for the mapping, and the total time elapsed was not included as an input. For model 2, only data from the current load pattern was used, but to capture long term trends it was necessary to also include the total time elapsed as an additional input.

Models 3 and 4 are analogous to models 1 and 2, except a linear kernel was used in the GP rather than a Mat\'ern kernel. This gives a simple base case for comparison. Using a linear kernel is equivalent to implementing Bayesian linear regression, and the key point to note in this context is that it provides far less flexibility for the model predictions compared with a Mat\'ern kernel.

Models 5 and 6 are also analogous to models 1 and 2, except that, rather than using a GP, they use a different regression technique called gradient boosting, as was introduced in section \ref{subsec:GB}.

\begin{figure*}[h]
%	\centering
%	{\includegraphics[width=0.75\textwidth]{\string"Figures/dQ_actual_vs_pred_all\string".pdf}}
	{\includegraphics[width=0.95\textwidth]{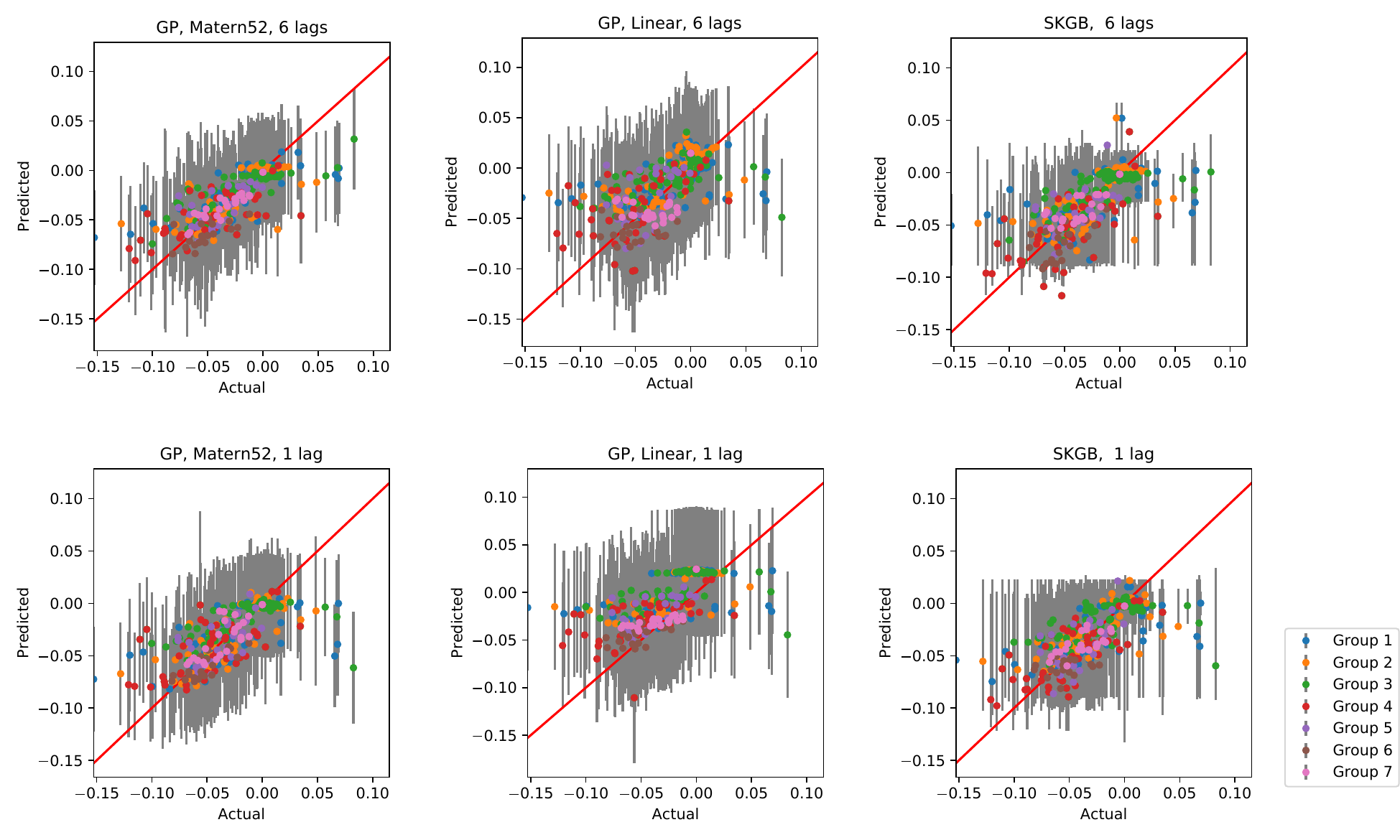}}
	\caption{Predicted versus actual $\Delta Q$ values for each dataset. The coloured markers indicate predicted values, and grey error bars indicate $\pm 2\sigma$ credibility intervals.}
	\label{fig:errors}
\end{figure*}

The predicted versus actual $\Delta Q$ for each approach is shown in Fig.\ \ref{fig:errors}. Model 1 was the best performing of the 6 cases tested, with $\text{RMSE}_{\Delta Q}$ and $\text{RMSE}_{Q}$ of $0.0201$ Ah and $0.07$ Ah respectively. Normalised capacity prediction error $\text{RMSE}_{\text{norm}}$ for model 1 was 4.3\%.

\begin{figure*}[h]
	\centering
	{\includegraphics[width=0.99\textwidth]{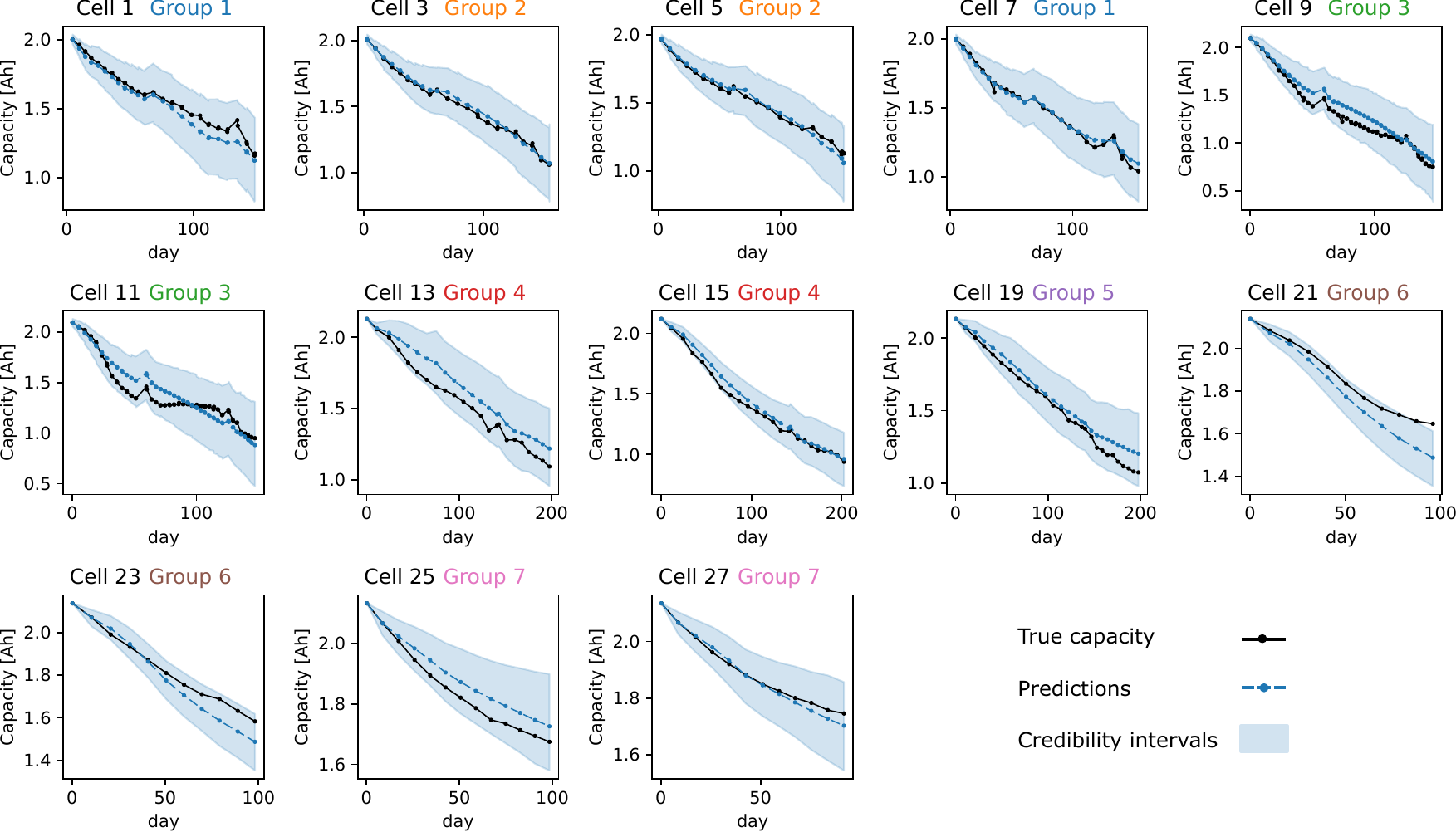}}
	\caption{Predicted capacity versus time on the test data. Black lines are true values, blue lines with markers are mean predictions and blue shaded region indicates $\pm 2\sigma$ credibility region.}
	\label{fig:example-results}
\end{figure*}

Finally we present in more detail in Fig.\ \ref{fig:example-results} the evolution of the capacity for each of the cells in the test dataset, using the best performing approach (model 1).

%\FloatBarrier
\section{Discussion\label{sec-discussion}}
The results given in section \ref{sec:results} show that model 1 accurately predicts the capacity trajectory, and provides reasonable, if slightly over-cautious, estimates of the uncertainty, indicated by the calibration score being close to 0.954. The true capacity generally lies within the $\pm 2\sigma$ interval denoted by the blue shaded region in Fig.~\ref{fig:example-results}.

The model is also seen to be capable of predicting both positive and negative capacity differences. For instance, it is apparent in Fig.~\ref{fig:example-results} that, although the capacities experience a long-term downward trend, they also experience occasional step increases.
The model correctly predicts the timing of a number of these instances, e.g.\ for cell 7 at day $\sim140$.
As an aside, the physical explanation for these increases is not clear; they may in fact be an artefact of the measurement process, possibly arising when reference tests are performed, after the cell is unused for some time. However, regardless of their cause, accounting for these effects is essential since the capacity measurement provided in a real application could also manifest similar behaviour.

Regarding feature selection, the fact that model 1 performs better than model 2 in the case of the dataset used here suggests that valuable information is being extracted from the inputs over the previous load patterns, which is not available from using just the total time elapsed as an additional input. 

Models 3 and 4, based on a linear kernel as noted earlier, perform considerably more poorly than the other approaches in terms of capacity prediction error, indicating that the simple linear combination of the inputs is insufficient to predict battery health for the dataset considered here, and the nonlinearities captured by the Mat\'ern kernel are significant in this case. Their calibration scores also indicate over-confidence.

The models based on gradient boosting are slightly less accurate in terms of mean predictions than models 1 and 2 and it is also noteworthy that they are erroneously over-confident, as indicated by their low calibration scores.

Finally, we note that the train/test split used in this paper (whereby the even numbered cells are used for training and the odd numbered cells for testing) ensures that there is at least one training cell in each of the 7 groups of differently cycled cells, Table \ref{tab:NASA-groups}. Inferior results may be obtained if this were not the case, e.g.\ if the first N cells were used for training and the remaining 26-N used for testing, since in the latter case the model would be extrapolating beyond the region of the input space used for training. In practice, the performance of these methods will rely on using a sufficiently large training set being available, such that a large range of input conditions are covered.

\section{Conclusions\label{sec-conclusions}}
This paper has developed a new technique for battery health prediction based on a Bayesian non-parametric model that estimates the change in capacity over a particular period of time as a function of how the battery was used during that period. A simple histogram-based feature selection approach was presented and models were trained using data from NASA \cite{bole2014randomized}. It was found that the best performing approach used Gaussian process regression with a Mat\'ern kernel function, and that time elapsed and charge throughput were the most important features to incorporate within the model, given the dataset used in this paper. It was also found that more accurate results could be achieved by considering the preceding 6 load patterns to capture longer range trends, rather than using absolute time as an input feature. Automated feature selection would be worth future investigation.

The best case results presented have a relative accuracy on mean capacity predictions that is within 5\% of the actual values. To our knowledge this is one of the first papers to actually quantify battery health predictive accuracy comprehensively, and this is one of the most accurate long range predictions of future capacity seen to date.

The approaches explored in this paper offer an interesting insight into how the stress factors that drive degradation actually influence the capacity trajectory. It is noteworthy that, despite having a dataset that includes a wide range of temperatures and currents, in this case it was found that time elapsed and charge throughput were the dominant inputs. However, a naive modelling approach that uses a simple linear combination of inputs results in very inaccurate predictions, as shown by the GP regression results using linear kernels.

There are a number of interesting next steps to explore. First, it would be useful to test these ideas against a much larger dataset to show their general validity and explore in more detail the sensitivity of the approach to additional inputs. Second, prior knowledge about expected degradation behaviour could be included as an extension to this work by including a parametric mean function within the GP framework. Third, in the present work, when the model is used predictively, it assumes perfect knowledge about the inputs, i.e.\ that the future current, voltage and temperature time series are known in advance. In practice this will not be the case, since depending on the application these variables depend on driving style or market conditions, ambient weather conditions etc. Predicting these inputs is a separate but important issue.
 
\section*{Acknowledgments}
This work was funded by Continental AG and an RCUK Engineering and Physical Sciences Research Council grant, ref.\ EP/K002252/1.

% References
%\FloatBarrier
\bibliographystyle{elsarticle-num}
\bibliography{GP_Paper}

\begin{thebibliography}{10}
\expandafter\ifx\csname url\endcsname\relax
  \def\url#1{\texttt{#1}}\fi
\expandafter\ifx\csname urlprefix\endcsname\relax\def\urlprefix{URL }\fi
\expandafter\ifx\csname href\endcsname\relax
  \def\href#1#2{#2} \def\path#1{#1}\fi

\bibitem{wankmueller2017impact}
F.~Wankmueller, P.~R. Thimmapuram, K.~G. Gallagher, A.~Botterud, {Impact of
  battery degradation on energy arbitrage revenue of grid-level energy
  storage}, Journal of Energy Storage 10 (2017) 56--66.

\bibitem{birkl2017degradation}
C.~R. Birkl, M.~R. Roberts, E.~McTurk, P.~G. Bruce, D.~A. Howey, {Degradation
  diagnostics for lithium ion cells}, Journal of Power Sources 341 (2017)
  373--386.

\bibitem{ruetschi2004aging}
P.~Ruetschi, Aging mechanisms and service life of lead--acid batteries, Journal
  of Power Sources 127~(1-2) (2004) 33--44.

\bibitem{farmann2015critical}
A.~Farmann, W.~Waag, A.~Marongiu, D.~U. Sauer, Critical review of on-board
  capacity estimation techniques for lithium-ion batteries in electric and
  hybrid electric vehicles, Journal of Power Sources 281 (2015) 114--130.

\bibitem{schimpe2018comprehensive}
M.~Schimpe, M.~von Kuepach, M.~Naumann, H.~Hesse, K.~Smith, A.~Jossen,
  Comprehensive modeling of temperature-dependent degradation mechanisms in
  lithium iron phosphate batteries, Journal of The Electrochemical Society
  165~(2) (2018) A181--A193.

\bibitem{wang2011cycle}
J.~Wang, P.~Liu, J.~Hicks-Garner, E.~Sherman, S.~Soukiazian, M.~Verbrugge,
  H.~Tataria, J.~Musser, P.~Finamore, Cycle-life model for
  graphite-{L}i{F}e{PO} 4 cells, Journal of Power Sources 196~(8) (2011)
  3942--3948.

\bibitem{schmalstieg2014holistic}
J.~Schmalstieg, S.~K{\"a}bitz, M.~Ecker, D.~U. Sauer, {A holistic aging model
  for Li (NiMnCo) O2 based 18650 lithium-ion batteries}, Journal of Power
  Sources 257 (2014) 325--334.

\bibitem{ecker2012development}
M.~Ecker, J.~B. Gerschler, J.~Vogel, S.~K{\"a}bitz, F.~Hust, P.~Dechent, D.~U.
  Sauer, Development of a lifetime prediction model for lithium-ion batteries
  based on extended accelerated aging test data, Journal of Power Sources 215
  (2012) 248--257.

\bibitem{dufo2014comparison}
R.~Dufo-L{\'o}pez, J.~M. Lujano-Rojas, J.~L. Bernal-Agust{\'\i}n, Comparison of
  different lead--acid battery lifetime prediction models for use in simulation
  of stand-alone photovoltaic systems, Applied Energy 115 (2014) 242--253.

\bibitem{birkl_howey_data}
C.~Birkl, D.~A. Howey, {Oxford Battery Degradation Dataset 1},
  \url{http://dx.doi.org/10.5287/bodleian:KO2kdmYGg} (2017).

\bibitem{harris2017failure}
S.~J. Harris, D.~J. Harris, C.~Li, Failure statistics for commercial lithium
  ion batteries: A study of 24 pouch cells, Journal of Power Sources 342 (2017)
  589--597.

\bibitem{kupper2017multi}
C.~Kupper, W.~G. Bessler, Multi-scale thermo-electrochemical modeling of
  performance and aging of a lifepo4/graphite lithium-ion cell, Journal of The
  Electrochemical Society 164~(2) (2017) A304--A320.

\bibitem{pinson2013theory}
M.~B. Pinson, M.~Z. Bazant, Theory of sei formation in rechargeable batteries:
  capacity fade, accelerated aging and lifetime prediction, Journal of the
  Electrochemical Society 160~(2) (2013) A243--A250.

\bibitem{Yang2017}
X.-G. Yang, Y.~Leng, G.~Zhang, S.~Ge, C.-Y. Wang,
  \href{http://dx.doi.org/10.1016/j.jpowsour.2017.05.110
  http://linkinghub.elsevier.com/retrieve/pii/S0378775317307619}{{Modeling of
  lithium plating induced aging of lithium-ion batteries: Transition from
  linear to nonlinear aging}}, Journal of Power Sources 360 (2017) 28--40.
\newline\urlprefix\url{http://dx.doi.org/10.1016/j.jpowsour.2017.05.110
  http://linkinghub.elsevier.com/retrieve/pii/S0378775317307619}

\bibitem{Deshpande2017}
R.~D. Deshpande, D.~M. Bernardi,
  \href{http://jes.ecsdl.org/lookup/doi/10.1149/2.0841702jes}{{Modeling
  Solid-Electrolyte Interphase (SEI) Fracture: Coupled Mechanical/Chemical
  Degradation of the Lithium Ion Battery}}, Journal of The Electrochemical
  Society 164~(2) (2017) A461--A474.
\newblock \href {http://dx.doi.org/10.1149/2.0841702jes}
  {\path{doi:10.1149/2.0841702jes}}.
\newline\urlprefix\url{http://jes.ecsdl.org/lookup/doi/10.1149/2.0841702jes}

\bibitem{hu2016battery}
X.~Hu, J.~Jiang, D.~Cao, B.~Egardt, Battery health prognosis for electric
  vehicles using sample entropy and sparse {B}ayesian predictive modeling, IEEE
  Transactions on Industrial Electronics 63~(4) (2016) 2645--2656.

\bibitem{patil2015novel}
M.~A. Patil, P.~Tagade, K.~S. Hariharan, S.~M. Kolake, T.~Song, T.~Yeo, S.~Doo,
  A novel multistage {S}upport {V}ector {M}achine based approach for li ion
  battery remaining useful life estimation, Applied Energy 159 (2015) 285--297.

\bibitem{wang2013prognostics}
D.~Wang, Q.~Miao, M.~Pecht, Prognostics of lithium-ion batteries based on
  relevance vectors and a conditional three-parameter capacity degradation
  model, Journal of Power Sources 239 (2013) 253--264.

\bibitem{nuhic2013health}
A.~Nuhic, T.~Terzimehic, T.~Soczka-Guth, M.~Buchholz, K.~Dietmayer, Health
  diagnosis and remaining useful life prognostics of lithium-ion batteries
  using data-driven methods, Journal of Power Sources 239 (2013) 680--688.

\bibitem{goebel2008prognostics}
K.~Goebel, B.~Saha, A.~Saxena, J.~R. Celaya, J.~P. Christophersen, Prognostics
  in battery health management, IEEE Instrumentation \& Measurement Magazine
  11~(4) (2008) 33.

\bibitem{saha2008uncertainty}
B.~Saha, K.~Goebel, Uncertainty management for diagnostics and prognostics of
  batteries using {B}ayesian techniques, in: Aerospace Conference, 2008 IEEE,
  IEEE, 2008, pp. 1--8.

\bibitem{he2011prognostics}
W.~He, N.~Williard, M.~Osterman, M.~Pecht, Prognostics of lithium-ion batteries
  based on {D}empster--{S}hafer theory and the {B}ayesian {M}onte {C}arlo
  method, Journal of Power Sources 196~(23) (2011) 10314--10321.

\bibitem{richardson2017gaussian}
R.~R. Richardson, M.~A. Osborne, D.~A. Howey, Gaussian process regression for
  forecasting battery state of health, Journal of Power Sources 357 (2017)
  209--219.

\bibitem{rasmussen2006gaussian}
C.~E. Rasmussen, Gaussian processes for machine learning, Citeseer, 2006.

\bibitem{murphy2012machine}
K.~P. Murphy, Machine learning: a probabilistic perspective, MIT press, 2012.

\bibitem{bishoppattern}
C.~M. Bishop, {Pattern Recognition and Machine Learning}, Springer, 2006.

\bibitem{bole2014randomized}
B.~Bole, C.~Kulkarni, M.~Daigle, Randomized battery usage data set, NASA AMES
  prognostics data repository.

\bibitem{bole2014adaptation}
B.~Bole, C.~S. Kulkarni, M.~Daigle, Adaptation of an electrochemistry-based
  li-ion battery model to account for deterioration observed under randomized
  use, in: Proceedings of Annual Conference of the Prognostics and Health
  Management Society, Fort Worth, TX, USA, Vol.~29, 2014.

\end{thebibliography}

\end{document}